
\input            harvmac
\input            epsf



\nref
\BergshoeffI
{
	E. Bergshoeff, E. Sezgin, and P. Townsend,
	``Supermembranes and Eleven Dimensional Supergravity,"
	Physics Letters Volume B189, ( Pg 75 )
}

\nref
\GromovI
{
	M. Gromov,
	``Pseudo Holomorphic Curves in Symplectic Manifolds,"
	Inventiones Mathematicae
	{\bf 82}, 307-347 (1985)
}

\nref
\HenneauxI
{
	M. Henneaux and C. Teitelboim,
	{\it Quantization of Gauge Systems } (1992)
	Princetion University Press,
	Princeton New Jersey.
}

\nref
\SingerI
{
	L. Baulieu and I. Singer,
	``The Topological Sigma Model,"
	Communications in Mathematical Physics
	{\bf 125} 227 (1989)
}

\nref
\StrommingerII
{
	K. Becker, M. Becker and A. Stromminger,
	``Fivebranes, Membranes and Non-Perturbative String Theory,"
	hep-th/9507158
}

\nref
\TownsendI
{
	M. Gunaydin, G. Sierra, and P. Townsend,
	``The Geometry of N = 2 Maxwell Einstein Supergravity
	and Jordan Algebras", Nuclear Physics B {\bf 242}
	(1984) 244
}
		
\nref
\WittenXI
{
	E. Witten,
	``Mirror Manifolds and Topological Field Theory,"
	hep-th/9112056; IASSNS-HEP-91/83;
	also in {\it Essays on Mirror Manifolds},
	ed. by S.T. Yau, International Press, 1992
}


\Title{
	\vbox{
		\baselineskip12pt
		\hbox{}
		\hbox{}
		\hbox{}
		}
	}
	{
	\vbox{
		\centerline{ GENERALIZED }
		\vskip2pt
		\centerline{ TOPOLOGICAL SIGMA MODEL }
		}
	}

\centerline{ Kelly Jay Davis }
\bigskip
\centerline{Rutgers University}
\centerline{Serin Physics Laboratory}
\centerline{Piscataway, NJ 08855}


\vskip .3in
\centerline{ABSTRACT}
In  this article we will examine a ``generalized topological
sigma model." This so-called ``generalized topological sigma
model" is the M-Theoretic analog of the standard topological
sigma model of string theory. We find that  the  observables
of the theory are elements in the cohomology  ring   of  the
moduli space of supersymmetric maps; in  addition,  we  find
that  the  correlation   functions   of   such   observables
allow  us  to compute non-perturbative  corrections  to  the
four-fermion   terms        present   in   M-Theory   on   a
six-dimensional Calabi-Yau.
\Date{2/25/97}


\newsec{ Introduction }

In this article we will derive the existence of, and examine
the properties of, a ``generalized topological sigma model."
This so-called ``generalized topological sigma model"  which
we will examine is the M-Theoretic analog  of  the  standard
topological sigma model encountered in string theory. So, in
particular, it is associated with a two-brane instead  of  a
one-brane as arises in the standard topological sigma model.
In addition, the  world-volume  action  of  our  generalized
topological sigma model is quite a bit different  from  that
encountered  in  the  standard  topological   sigma   model.
However, we will find that many of the generic properties of
the generalized topological sigma model are very similar  to
those  of  the  standard  topological sigma  model,  and  it
is  these  similarities  which  will  allow  us  to   better
understand  the properties of  the  generalized  topological
sigma  model  and  the  relation  of  these  properties   to
M-Theory.

As one will recall \WittenXI, the standard topological sigma
model is a two-dimensional field theory  which describes the
dynamics of a map from the two-dimensional world-sheet to  a
six-dimensional target space. This model can  be  derived by
two different, yet equivalent   methods.  The  first  is  by
``twisting" the standard sigma model \WittenXI.  The  second
is by simply  postulating  a  two-dimensional  action,  then
gauge fixing this action to prove it is  equivalent  to  the
above ``twisted" action \SingerI. Both of these methods have
their strengths and weaknesses. However, we will  rely  only
upon this second method as its  analog  will  occur  in  the
generalized context with which we are concerned.

Upon deriving the action of the standard  topological  sigma
model, as one will  recall  \WittenXI,  it  is  possible  to
employ this action to calculate the Yukawa couplings of  the
full, ``un-twisted" string theory.  This   is  a  relatively
novel result. One can calculate a  complicated  result,  the
Yukawa  couplings,  in  a  string theory  by   examining   a
marginally related theory, the  standard  topological  sigma
model. In addition, these string theoretic  results  tie  in
very nicely with Gromov theory  \GromovI,  and  both  string
theory and Gromov theory benefit from this  interaction.  At
best, we could hope that the generalized  topological  sigma
model, which we examine here, will bear similar fruit.

As we remarked  earlier,  the  action  for  our  generalized
topological sigma model will arise by  way  of  an  educated
guess. More to the point, this  educated  guess  will  arise
upon   closely   examining   the    low-energy     effective
world-volume  action  of  the  M-Theoretic  two-brane.  This
action admits certain field  configurations  which  globally
minimize the action, so-called {\it instantons}. Furthermore,
the  lower  bound  on  this  action  is   provided   by   an
``action-like" integral. It is this ``action-like"  integral
which we  shall take as the  action  for   our   generalized
topological  sigma model. This action selection  process  is
not as strange as it  first  may  seem.  A  similar  process
applied  to  the  standard sigma model leads to the standard
topological  sigma  model action. In addition, applying this
process to the Yang-Mills action  leads  to  the  action for
topological  Yang-Mills theory. So,  in  defining the action
of  the  generalized topological sigma model in this manner,
we  are  actually in good company.

Our next step is to then examine the generalized topological
sigma model action in a bit more detail. Our first  step  in
this direction is to gauge fix the action  obtained  by  the
above action selection process. To achieve this goal we will
employ the now standard BV method of gauge fixing \HenneauxI.
This will provide us with  a new, gauge fixed version of the
generalized topological sigma model action. It is from  this
gauge  fixed  action  that   we   may   derive   ``physical"
implications. We will find that  the  observables associated
with this gauge fixed action arise in a manner very  similar
to the observables of the standard topological sigma  model.
In addition, we will also find that the observables  of  the
generalized  topological  sigma   model   have   correlation
functions which may be computed in a manner very similar  to
that encountered in the  standard  topological  sigma model.
Finally,  upon examining  how  to  compute  the  correlation
functions  of  various  observables   in   the   generalized
topological  sigma model, we  will examine  the  ``physical"
relevance  of such correlation functions to five dimensional
M-Theory.


\newsec{ Generalized Topological Sigma Model }

In this section we will derive  the  existence  of, and  the
properties of, the generalized topological sigma model.  Our
first  goal, which we pursue in the next subsection,  is  to
derive  the action  of  the  generalized  topological  sigma
model. After identifying the action of this  model, we  will
then proceed to examine various properties of  this  action.
So, let us now begin  our  examination  of  the  generalized
topological  sigma model.


\subsec{ Generalized Topological Sigma Model Action }

In  this  subsection  we  will  derive  the  action  of  the
generalized topological sigma model. Our plan of  ``action,"
as outlined in the  introduction, is  as  follows:  We  will
examine the low-energy effective world-volume action of  the
M-Theoretic two-brane to  determine  a  lower-bound  on  the
value of this action. This lower-bound will  appear  in  the
form of an ``action-like" integral. This  integral  will  be
our action for this generalized topological sigma model. Let
us now begin with this derivation.

Consider the two-brane of M-Theory. By way of early  results
in  eleven-dimensional  supergravity  \BergshoeffI,  we  may
write down the low-energy  effective world-volume action  of
this two-brane. It is \StrommingerII,

\eqnn\LowEnergyTwoBraneActionI
$$
\eqalignno{
	S =
	\left(
		{ {1} \over L_{P}^{3} }
	\right)
	\int_{ \Sigma_{3} }
	d^{3}\sigma \sqrt{ h }
	\biggl(
		&{1 \over 2}
		h^{\alpha \beta}
		\partial_{\alpha}X^{M}\partial_{\beta}X^{N}
		g_{MN} -
		{1 \over 2}
		-{i {\overline \Theta} \Gamma^{\alpha}
		\nabla_{\alpha} \Theta}
		&\LowEnergyTwoBraneActionI \cr
		&+{{i \over {3!}}
		\epsilon^{\alpha\beta\gamma}A_{MNP}
		\partial_{\alpha}X^{M}\partial_{\beta}X^{N}
		\partial_{\gamma}X^{P} +
		\cdots
	\biggr),}\cr
}
$$

\noindent where $L_{p}$ is  the  eleven-dimensional  Plank's
length, $\Sigma_{3}$ the world-volume, $h_{\alpha\beta}$ the
world-volume metric, $X^{M}$ a map from $\Sigma_{3}$ to  the
eleven-dimensional target, $g_{MN}$  the  eleven-dimensional
metric,  $\Theta$  an   eleven-dimensional   Driac   spinor,
$\Gamma^{M}$   eleven-dimensional   Driac   matrices,    and
$A_{MNP}$ is the M-Theoretic three-form. This action  admits
two-different fermionic symmetries. The first of  these  two
symmetries  is  given  by  the   following   transformations
\StrommingerII,

\eqn
\GlobalFermionicSymmetryI{
	\eqalign{
		\delta_{\epsilon}\Theta &= \epsilon, \cr
		\delta_{\epsilon}X^{M}  &=
		 i{\overline\epsilon}\Gamma^{M}\Theta,\cr
	}
}

\noindent where $\epsilon$ is a constant  eleven-dimensional
Driac spinor. This symmetry is simply a  reflection  of  the
fact  that the target space theory, M-Theory in  this  case,
is supersymmetric. The second of these two symmetries is the
so-called $\kappa$-symmetry. It is defined by the  following
transformations \StrommingerII,

\eqn
\LocalFermionicSymmetryI{
	\eqalign{
	\delta_{\kappa}\Theta &=
		2P_{+}\kappa( \sigma ), \cr
	\delta_{\kappa}X^{M}  &=
		2i{\overline \Theta}
		\Gamma^{M}P_{+}\kappa ( \sigma ), \cr
	}
}

\noindent where $\kappa( \sigma )$ is an  eleven-dimensional
Driac  spinor  which  is  not  necessarily   constant   and
\StrommingerII\

\eqn
\ProjectionOperatorsI{
	P_{\pm} =
	{ 1 \over 2 }
	\left(
		1 \pm
		{ i \over {3!} }
		\epsilon^{\alpha\beta\gamma}
		\partial_{\alpha}X^{M}
		\partial_{\beta}X^{N}\partial_{\gamma}X^{P}
		\Gamma_{MNP}
	\right)
}

\noindent are projection operators.  Now,  the  question  we
wish to answer about this action  is:  What  are  the  field
configurations which minimize the action?  In  other  words,
what are the instantons of this theory?

The trivial answer to this question is  relatively  easy  to
come by. The field configuration $X^{M}(\sigma) = x^{M}$ and
$\Theta   =    0$      has      a      vanishing      action
and   is    a    minimum.    However,    as     we     shall
see, there also  exist  non-trivial  minima  of  the  action
$S$. These may be found by  looking  a  bit  closer  at  the
action  $S$.  The  first  two  terms  of  the   action   $S$
classically reduce  to  the volume of  the  world-volume.  In
addition, as any  minimum  of  the  action  is  a  classical
solution, for our minima we may assume that  the  first  two
terms of the action $S$  simply  yield  the  volume  of  the
world-volume.  Also,  the  fourth   term   in   the   action
may be written as follows,

\eqn
\ThirdActionTerm{
	{ {1} \over {3!} }
	\int_{\Sigma_{3}}
		d^{3}\sigma \sqrt{h} 
		\left(
			\epsilon^{\alpha\beta\gamma}
			\partial_{\alpha}X^{M}
			\partial_{\beta}X^{N}
			\partial_{\gamma}X^{P}
			A_{MNP}
		\right) =
	\int_{\Sigma_{3}}
		X^{*}( A ),
}

\noindent where $X^{*}(A)$ is the pull-back to  $\Sigma_{3}$
of $A$ by way of $X$. This implies  that  this  fourth  term
only depends upon the homotopy class of X and the cohomology
class of\foot{ Note, \StrommingerII\ $A$ must be  closed  to
define a supersymmetric compactification of M-Theory, and we
will assume that $A$ is indeed closed. } $A$.  Hence, within
a given homotopy class $[X]$ and for a fixed $A$, the fourth
term in the action is independent of the $X$  we  choose  to
represent $[X]$. The  remaining  terms  in  the  action  all
contain fermions; thus, we can, for the moment, ignore  them
for our present purpose. So, from the above arguments we may
see that to minimize $S$ in a  given  homotopy  class  $[X]$
we must minimize the volume of $\Sigma_3$. Now, for a  given
$[X]$ let us examine what conditions are  imposed  upon  $X$
representing $[X]$ so  that  $X$  minimizes  the  volume  of
$\Sigma_{3}$.

As we wish to examine, in addition to the  trivial  case  of
$X^{M}(\sigma) = x^{M}$,  non-trivial  field  configurations
$X^{M}(\sigma) \ne x^{M}$, we must establish some  means  of
obtaining non-trivial homotopy classes $[X]$.  This  can  be
done by giving the two-brane $\Sigma_{3}$ something to  wrap
about. We will achieve this by assuming the target  manifold
$X_{11}$ is of the following form,

\eqn
\TargetManifoldSplit{
	X_{11} \cong M_{5} \times X_{6},
}

\noindent where $M_{5}$ is  a  Minkowski  five-manifold  and
$X_{6}$ is a six-dimensional Calabi-Yau. If we take $X_{11}$
to be of this form, then we may wrap $\Sigma_{3}$ completely
about the homology of $X_6$ to obtain  non-trivial  homotopy
classes $[X]$. With  this  form  for  $X_{11}$, let  us  now
consider what conditions are imposed upon   $X$ representing
$[X]$ so  that  it  minimizes  the volume of $\Sigma_{3}$.

To obtain the constraints imposed upon a $X$ which  minimize
the volume of $\Sigma_{3}$ we will  consider  the  following
inequality,

\eqn
\VolumeBoundIntegralI{
	\int_{\Sigma_{3}} d^{3}\sigma \sqrt{h} \,\,
		\left(
			{\overline \epsilon}_{\theta}
			P^{\dagger}_{-}P_{-}
			\epsilon_{\theta}
		\right)  \ge 0
}

\noindent where

\eqn
\ThetaSpinorDefinition{
	\epsilon_{\theta} =
	\left(
		e^{i\theta} \epsilon_{+} +
		e^{-i\theta} \epsilon_{-}
	\right)	
}

\noindent  and  $\epsilon_{\pm}$  are  covariantly  constant
spinors of opposite chirality on $X_{6}$  which  are complex
conjugates of one another  $\epsilon_{+}=(\epsilon_{-})^{*}$
and  $\theta  \in  {\bf R}$.   One  can   always   normalize
$\epsilon_{\pm}$ such that \StrommingerII,

\eqn\SpinorNormalization{
	\eqalign{
	  \gamma_{mnp} \epsilon_{+} &=
	    e^{-{\cal K}} \Omega_{mnp} \epsilon_{-},\cr
	  \gamma_{{\bar m} np} \epsilon_{+} &=
	   2iJ_{{\bar m}[n}\gamma_{p]} \epsilon_{+},\cr
	  \gamma_{{\bar m}} \epsilon_+ &= 0,\cr
	}
}

\noindent where $\Omega$ is the  holomorphic  three-form  on
$X_{6}$, $J$ is the Kahler form on $X_{6}$, $\gamma_{m}$ and
$\gamma_{\overline m}$ are gamma matrices on  $X_{6}$,  and
${\cal K}$ is a function given by,

\eqn
\KDefinitionI{
	{\cal K}=
	{1\over 2}
	\left(
		{\cal K}_{\cal V}-
		{\cal K}_{\cal H}
	\right),
}

\noindent where

\eqn
\KDefinitionII{
	{\cal K}_{\cal H}=
	-\log
	\left(
		i \int_{X_{6}}
			\Omega \wedge {\overline \Omega}
	\right) 
}

\noindent and

\eqn
\KDefinitionIII{
	{\cal K}_{\cal V}=
	-\log
	\left(
		{4 \over 3} \int_{X_{6}}
			J \wedge J \wedge J
	\right).
}

\noindent These normalization conditions and  our inequality
imply,

\eqn
\VolumeBoundIntegralII{
	V(\Sigma_{3}) \ge
	{1 \over 2} e^{-{\cal K}}
	\left(
	 e^{ i\phi} \int_{\Sigma_{3}} X^{*}(\Omega) +
	 e^{-i\phi} \int_{\Sigma_{3}} X^{*}({\overline \Omega})
	\right),
}

\noindent where $\phi = 2\theta+\pi/ 2$ and  $V(\Sigma_{3})$
is the volume of $\Sigma_{3}$. Adjusting $\phi$ to  maximize
the right-hand side, one has,

\eqn
\VolumeBoundIntegralIII{
	V(\Sigma_{3}) \ge
	e^{-{\cal K}}
	\left|
		\int_{\Sigma_{3}}
			X^{*}(\Omega)
	\right| .
}

\noindent So, the volume $V(\Sigma_{3})$ of $\Sigma_{3}$  is
bounded below by the above integral.  This, in turn, implies
that the action $S$ is bounded below by,

\eqn
\ActionBoundIntegralI{
	| S | \ge
	\left(
		{ {1} \over {L^{3}_{P}} }
	\right)
	\left(
		e^{-{\cal K}}
		\left|
			\int_{\Sigma_{3}}
			X^{*}(\Omega)
		\right| +
		\int_{\Sigma_{3}} X^{*}(A)
	\right).
}

\noindent Furthermore, one may see, by way of  our  original
inequality     \VolumeBoundIntegralI,     the     inequality
\ActionBoundIntegralI\ is saturated if, and only if,

\eqn
\BPSConditionI{
	P_{-} \epsilon_{\theta} = 0.
}

\noindent  In  other  words,  instantons  of  the  two-brane
world-volume satisfy  $P_{-}\epsilon_{\theta} = 0$.  As  was
proven previously \StrommingerII, this occurs if,  and  only
if, the following conditions are satisfied,

\eqn
\BPSConditionII{
	X^{*}(J) = 0
}

\noindent and

\eqn
\BPSConditionIII{
	X^{*}(\Omega) =
	e^{-i\phi}e^{\cal K} \epsilon,
}

\noindent  where  $\epsilon$  is   a   volume   element   on
$\Sigma_{3}$. Now, as we mentioned earlier, we will use this
``action-like" lower bound as the action for the generalized
topological sigma model. So, with this in  mind,  we  define
the generalized topological sigma model as the  world-volume
theory with the following action,

\eqn
\TopologicalSigmaModelActionI{
	S_{0} =
	\left(
		{ {1} \over {L^{3}_{P}} }
	\right)
	\left(
		e^{-{\cal K}}
		\left|
			\int_{\Sigma_{3}}
				X^{*}(\Omega)
		\right| +
		i \int_{\Sigma_{3}} X^{*}(A)
	\right).
}

\noindent Next, let us examine in detail the  properties  of
the theory defined by the above action.


\subsec{ Generalized Topological Sigma Model }

In this subsection we will examine in detail the  properties
of the theory defined by the generalized  topological  sigma
model action,

\eqn
\TopologicalSigmaModelActionII{
	S_{0} =
	\left(
		{ {1} \over {L^{3}_{P}} }
	\right)
	\left(
		e^{-{\cal K}}
		\left|
			\int_{\Sigma_{3}}
				X^{*}(\Omega)
		\right| +
		i \int_{\Sigma_{3}} X^{*}(A)
	\right) .
}

\noindent So, let us begin.


\vskip .5in
\noindent {\it 2.2.1 Gauge Fixing }
\vskip .25in

First, we will examine  why  such  an  action  is considered
``topological." This is rather easy to understand.  Consider
the generalized topological sigma model action above. Rather
obviously, none of the  terms  in  the  action  involve  the
metric $h_{\alpha\beta}$ on $\Sigma_3$. So, the action,  and
the  resultant  theory\foot  { Mod   any   Chern-Simons-like
``framing" effects.}, are  independent  of  the world-volume
metric. In addition  to  this  ``topological"  property, the
action also admits a ``large" symmetry group involving  $X$.
The generalized topological sigma model action  consists  of
various   integrals   with   integrands    of    the    form
$X^{*} (\cdots)$, where $`` \cdots "$ is some closed form on
$X_{6}$.  So,  this  implies  that  the  action,  and   thus
the resultant theory, only depend upon  the  homotopy  class
$[X]$ of the map $X$. This, in turn, implies that the action
of the generalized  topological  sigma  model  is  invariant
under arbitrary infinitesimal variations  of  $X$.  This  is
indeed a ``large" gauge symmetry, as advertised.

Out next goal is to gauge fix  the  generalized  topological
sigma model action. In gauge  fixing  this  action, we  will
employ  the  now  standard  BV   gauge   fixing   procedure
\HenneauxI. The first step in this process is  to  introduce
the correct field/anti-field content; we now  proceed  with
this step. First, one introduces a Grassmann  odd anti-field
$X^{*}_{3i}$, where  $i \in \{ 1, \dots , 6 \}$,  at   ghost
number $-1$, associated with  the  field  $X^{i}$.  We  will
take  $X^{*}_{3i}$  to  be   a   section   of   $\Lambda^{3}
(\Sigma_{3})$;  note,   normally   $X^{*}_{3i}$   would   be
taken  to  be  a  section  of $\Lambda^0 (\Sigma_{3})$,  but
we can take it to be a section of  $\Lambda^{3}(\Sigma_{3})$
by  way  of  duality.  We  next  introduce  the  fields  and
anti-fields associated with the gauge symmetries of $S_{0}$.
As the action $S_0$ is invariant with respect  to  arbitrary
infinitesimal variations of $X$,

\eqn
\XGaugeSymmetry{
	\delta X^{i} = \delta^{i}_{j} \epsilon^{j},
}

\noindent where $\delta^{i}_{j}$ is  a  delta  function  and
$\epsilon^{i}$ is an arbitrary infinitesimal  parameter,  we
must introduce a Grassmann odd field $\chi^{i}$, a   section
of $X^{*}( TX_{6} )$, at ghost number $1$. In addition,  one
should    introduce     the     corresponding     anti-field
$\chi^{*}_{3i}$,  a  section  of  $X^{*}(\Lambda^{1}(X_{6}))
\otimes    \Lambda^{ 3 } ( \Sigma_{ 3 } )$,     which     is
Grassmann even and at ghost number $-2$. Now, as  the  delta
function $\delta^{i}_j$ appears in equation \XGaugeSymmetry\
and an arbitrary  $X$  solves  the  classical  equations  of
motion\foot {$S_{0}$ only depends upon $[X]$.},  the  theory
is irreducible. Hence, one does not need  to  introduce  any
more fields or anti-fields. However, one could, and we will,
introduce extra, non-minimal fields and anti-fields. Let  us
now define these non-minimal fields and anti-fields.

Generically, non-minimal  fields  and  anti-fields  must  be
introduced  in  groups  of  four.  So,  for   example,   one
introduces the fields and anti-fields $A$, $A^{*}$, $B$, and
$B^{*}$ such that,

\eqn
\GenericNonminimalFields{
	\eqalign{
		gh( A ) + gh( A^{*} ) &=  - 1 \cr
		gh( B ) + gh( B^{*} ) &=  - 1 \cr
		gh( A ) - gh( B^{\,\,} ) &=  - 1, \cr
	}
}

\noindent   where   $gh(\cdots)$   denotes   ghost   number.
Specifically, we will introduce two  such  groups  of  four.
The first group of four consists of two fields  $b_{0}$  and
$C_{0}$ which are world-volume scalars. $C_{0}$ is Grassmann
odd and at ghost number $-1$ while $b_{0}$ is Grassmann even
and at ghost number $0$.  The  anti-field  corresponding  to
$C_{0}$  is  $C^{*}_{3}$, a  world-volume   three-form\foot{
Normally, $C^{*}_{3}$ would be taken to  be  a  world-volume
scalar, but via duality we may consider  it  as  three-form.
Similar arguments also apply for subsequent field/anti-field
pairs.}; $C^{*}_3$ is Grassmann even at ghost number 0.  The
anti-field  corresponding  to  $b_{0}$  is   $b^{*}_{3}$,  a
world-volume three-form. $b^{*}_{3}$  is  Grassmann  odd  at
ghost number -1. Our second group of four is similar to  the
first. The fields are $E_1$, a one-form at ghost number  -1,
and $d_1$, a one-form at ghost number 0. The anti-fields are
$E^{*}_{2}$, a two-form at ghost number $0$, and  $d^{*}_2$,
a two-form at ghost number -1. Also,  all fields/anti-fields
in this group with even ghost number are Grassmann  even and
all those with odd ghost number are Grassmann odd.

Now, let  us  find the ``quantum action." This is relatively
easy to do in this case. One  can  first  solve  the  master
equation to obtain,

\eqn
\MasterEquationSolution{
	S =
 		S_{0} +
 		\left(
 			{ {1} \over {L^{3}_{P}} }
 		\right)
 		\int_{\Sigma_{3}}
 			C^{*}_{3}   \wedge   b_{0}  +
 			E^{*}_{2}   \wedge   d_{1}  +
 			X^{*}_{3i}  \wedge  \chi^{i}.
}

\noindent This proper solution to the master equation, as it
turns  out,  is  also   a   solution  to  the quantum master
equation. So, our quantum action $W$ is given by,

\eqn
\QuantumMasterEquationSolution{
	\eqalign{
		W =
 			\left(
 				{ {1} \over {L^{3}_{P}} }
			\right)
			\Biggl(
				e^{-{\cal K}}
				\left|
				\int_{\Sigma_{3}}
					X^{*}(\Omega)
				\right| &+
				i \int_{\Sigma_{3}} X^{*}(A)
			\Biggr) + \cr
 			&\left(
 				{ {1} \over {L^{3}_{P}} }
 			\right)
 			\int_{\Sigma_{3}}
 				C^{*}_{3}   \wedge   b_{0}  +
 				E^{*}_{2}   \wedge   d_{1}  +
 				X^{*}_{3i}  \wedge  \chi^{i}.
 	}
}

\noindent Now, to fix the  final  remaining  gauge  symmetry
present in $W$ we  must  introduce  a  Grassmann  odd  gauge
fixing fermion,

\eqn
\GaugeFixingFermionI{
	\psi =
	\psi
	\left(
		X^{i},  \chi^{i},
		b_{0},  C_{0},
		d_{1},  E_{1}
	\right),
}

\noindent at ghost  number  $-1$.  With  this  gauge  fixing
fermion, one gauge fixes the quantum action $W$ as follows,

\eqn
\QMESGaugeFixedI{
	\eqalign{
		W_{\psi} =
 			\left(
 				{ {1} \over {L^{3}_{P}} }
			\right)
			\Biggl(
				e^{-{\cal K}}
				&\left|
				\int_{\Sigma_{3}}
					X^{*}(\Omega)
				\right| +
				i \int_{\Sigma_{3}} X^{*}(A)
			\Biggr) + \cr
 			& \left(
 				{ {1} \over {L^{3}_{P}} }
 			\right)
 			\int_{\Sigma_{3}}
 			\left(
 				{ {\delta\psi} \over {\delta C_{0}}  }
 				\wedge   b_{0}  +
 				{ {\delta\psi} \over {\delta E_{1}}  }
 				\wedge   d_{1}  +
 				{ {\delta\psi} \over {\delta X^{i}}  }
 				\wedge \chi^{i}
 			\right) .
 	}
}

\noindent Now, the question is, what gauge fixing fermion to
choose? This is the question we will answer next.

To fix the final remaining gauge symmetry one must choose  a
gauge  fixing  fermion  $\psi$  such  that  it   defines   a
Lagrangian submanifold of ``field space."  In  other  words,
one must choose $\psi$ such that the zeros of the functions

\eqn
\AlgebraicLagrangian{
	\Omega_{A} \equiv
	\left(
		\Phi^{*}_{A} -
		\left(
			{ {\delta\psi} \over {\delta\Phi^{A}} }
		\right)
	\right),
}

\noindent  where  $\Phi^{A}$  is   a   generic   field   and
$\Phi^{*}_{A}$   its   anti-field,   define   a   Lagrangian
submanifold  of  ``field  space."   This   occurs   if   the
classical anti-bracket  of  $\Omega_{A}$  with  $\Omega_{B}$
vanishes. One can see that this  occurs  for  the  following
choice of $\psi$,

\eqn
\GaugeFixingFermionII{
	\psi =
	\int_{\Sigma_{3}}
		C_{0} \wedge
		\left(
			X^{*}(\Omega) - e^{-i\phi}e^{\cal K} \epsilon
		\right) +
		E_{1} \wedge
		X^{*}(J),
}

\noindent where $\epsilon$ is a volume-element on $\Sigma_3$.
This gauge fixing fermion leads to the following gauge fixed
quantum action,

\eqn
\QMESGaugeFixedII{
	\eqalign{
		W_{\psi} =
 			&\left(
 				{ {1} \over {L^{3}_{P}} }
			\right)
			\Biggl(
				e^{-{\cal K}}
				\left|
				\int_{\Sigma_{3}}
					X^{*}(\Omega)
				\right| +
				i \int_{\Sigma_{3}} X^{*}(A)
			\Biggr) + \cr
 			& \left(
 				{ {1} \over {L^{3}_{P}} }
 			\right)
 			\int_{\Sigma_{3}}
 			\Biggl(
 			 \left(
				X^{*}(\Omega) - e^{-i\phi}e^{\cal K}\epsilon
			 \right)
 			 \wedge   b_{0}  +
 			 X^{*}(J)
 			 \wedge   d_{1}
 			 \Biggr) + \cr
 			 &\left(
 				{ {1} \over {3! L^{3}_{P}} }
			\right)
			\int_{\Sigma_{3}}
			 \bigg(
			 	C_{0}
			 	D_{\alpha}\chi^{m}
			 	\partial_{\beta} X^{n}
			 	\partial_{\gamma}X^{p}
			 	\Omega_{mnp} +
			 	C_{0}
			 	\partial_{\alpha}X^{m}
			 	D_{\beta}\chi^{n}
			 	\partial_{\gamma}X^{p}
			 	\Omega_{mnp} + \cr
			 	& \qquad\qquad\quad\quad\,\,\, C_{0}
			 	\partial_{\alpha}X^{m}
			 	\partial_{\beta} X^{n}
			 	D_{\gamma}\chi^{p}
			 	\Omega_{mnp} +
			 	C_{0}
			 	\partial_{\alpha}X^{m}
			 	\partial_{\beta} X^{n}
			 	\partial_{\gamma}X^{p}
			 	\partial_{i}\Omega_{mnp}
			 	\chi^{i}     - \cr
			 	& \qquad\qquad\quad\quad\,\,\, { 1 \over 2 }
			 	e^{-i\phi} e^{\cal K} C_{0}
			 	\left(
			 	Det
			 	\partial_{\alpha} X^{i}
			 	\partial_{\beta}  X^{j}
			 	g_{ij}
			 	\right)^{1/2}
			 	\partial^{\alpha} X^{i}
			 	\partial^{\beta}  X^{j}
			 	g_{ij} \cr
			 	& \qquad\qquad\quad\quad\,\,\,
			 	\bigg(
			 	D_{\alpha}       \chi^{i}
			 	\partial_{\beta} X^{j}
			 	g_{ij} +
			 	\partial_{\alpha} X^{i}
			 	D_{\beta}        \chi^{j}
			 	g_{ij} +
			 	\partial_{\alpha} X^{i}
			 	\partial_{\beta}  X^{j}
			 	\partial_{k} g_{ij} \chi^{k}
			 	\bigg)
			 	\epsilon_{\alpha\beta\gamma} + \cr
			 	& \qquad\qquad\quad\quad\,\,\, 3 E_{1\alpha}
			 	D_{\beta}\chi^{m}
			 	\partial_{\gamma}X^{\overline n}
			 	J_{m{\overline n}} +
			 	3 E_{1\alpha}
			 	\partial_{\beta} X^{m}
			 	D_{\gamma}\chi^{\overline n}
			 	J_{m{\overline n}} + \cr
			 	& \qquad\qquad\quad\quad\,\,\, 3 E_{1\alpha}
			 	\partial_{\beta} X^{m}
			 	\partial_{\gamma}X^{\overline n}
			 	\partial_{i}
			 	J_{m{\overline n}}
			 	\chi^{i}
			 \bigg)
			 d\sigma^{\alpha} \wedge
			 d\sigma^{\beta}  \wedge
			 d\sigma^{\gamma},
 	}
}

\noindent where $D$ is the properly twisted version of  $d$,
$g_{ij}$ is the metric on $X_{6}$, and $\epsilon_{123} = 1$.
Now, let  us  examine this gauge fixed action.


\vskip .25in
\noindent {\it 2.2.2 Geometry of $W_{\psi}$ }
\vskip .25in

Upon looking  at  the  above  gauge  fixed  action, one  may
ascertain  various  interesting  facts.  First  of  all, the
equation of motion for $b_{0}$ is 

\eqn
\PathIntegralConstraintI{
	X^{*}(\Omega) - e^{-i\phi}e^{\cal K}\epsilon = 0
}

\noindent and the equation of motion for $d_{1}$ is

\eqn
\PathIntegralConstraintII{
	X^{*}(J) = 0.
}

\noindent Now this is  rather interesting, as the two  above
constraints  are   the   same constraints   we   encountered
earlier  when   we    were characterizing instantons for the
field  $X$.  So,  in  other  words,  we  have   found   that
classically $X$ is an instanton in this theory. Now, looking
at  \QMESGaugeFixedII\  the  question  may  arise,  what  of
$\chi^i$, $C_{0}$, and $E_{1}$. These fields,  however, take
a bit more work than $X$.

As one can tell from the above quantum action, the  equation
of  motion  for  $C_{0}$  enforces   the   constraint   that
$\chi^{i}$ lie in the cotangent space of an  $X$  satisfying
equation \PathIntegralConstraintI. Similarly,  the  equation
of  motion  for  $E_{1}$  enforces   the   constraint   that
$\chi^{i}$ lie in the cotangent space of  a  $X$  satisfying
equation \PathIntegralConstraintII. Hence, the equations  of
motion  for  $C_{0}$  and  $E_{1}$   together   imply   that
classically $\chi^{i}$ is an element of the cotangent  space
of instanton  moduli  space.  Next,  let  us  see  what  the
``topological" properties of this  action  imply  about  the
above equations.

Looking at equation \QMESGaugeFixedI\ one can  see  that  it
may be written in the following manner,

\eqn
\QMESGaugeFixedIV{
		W_{\psi} =
 			\left(
 				{ {1} \over {L^{3}_{P}} }
			\right)
			\left(
				e^{-{\cal K}}
				\left|
				\int_{\Sigma_{3}}
					X^{*}(\Omega)
				\right| +
				i \int_{\Sigma_{3}} X^{*}(A) +
 				L^{3}_{P} \int_{\Sigma_{3}}
 					Q \psi
 			\right),
}

\noindent where $Q$ is the BRST operator. Now, as the theory
is independent of which gauge fixing  fermion  one  chooses,
we are free to choose instead of our  gauge  fixing  fermion
$\psi$ the gauge fixing fermion  $t\psi$,  where  $t$  is  a
constant. This implies that the  quantum  action  takes  the
form,

\eqn
\QMESGaugeFixedV{
		W_{\psi} =
 			\left(
 				{ {1} \over {L^{3}_{P}} }
			\right)
			\left(
				e^{-{\cal K}}
				\left|
				\int_{\Sigma_{3}}
					X^{*}(\Omega)
				\right| +
				i \int_{\Sigma_{3}} X^{*}(A) +
 				t L^{3}_{P} \int_{\Sigma_{3}}
 					Q \psi
 			\right).
}

\noindent However, the theory is independent of  the  actual
value of $t$ as it is  independent  of  which  gauge  fixing
fermion one chooses. Thus, we are free to take $t\rightarrow
\infty$. Doing  so,  we  find  that  the   above   equations
\PathIntegralConstraintI\   and   \PathIntegralConstraintII\
along with the classical fact that $\chi^{i}$ is an  element
of the cotangent space of instanton  moduli  space  all hold
exactly in the quantum theory. Hence, $\chi^{i}$ in the path
integral actually  defines  an  element  of   the  cotangent
space  of  instanton  moduli  space  and  $X$  is actually an
instanton due  to  this localization.


\vskip .25in
\noindent {\it 2.2.3 Observables of $W_{\psi}$ }
\vskip .25in

Let us  now  examine the  observables  of  this  generalized
topological  sigma  model.  As  is  well  known  \HenneauxI,
observables of a BV gauge fixed theory are the BRST operator
cohomology classes. In  other  words,  an  observable  is  a
function of the theory's fields which is $Q$ closed, but not
$Q$ exact. In our particular case, it is possible to  obtain
a rich geometrical interpretation of such observables.

To do so, let us first note the action of $Q$ on $X$. By way
of the definition of $Q$, one has,

\eqn
\QActionOnX{
	\eqalign{
		QX &= ( X, W ) + \Delta X				\cr
		   &= ( X, W )							\cr
		   &= \left(
		   	  { {\chi} \over {L^{3}_{P}} }
		   	  \right),	\cr
	}
}

\noindent where $(\,\, , \,\,)$ denotes the anti-bracket and
$\Delta$ is the standard ``Delta" operator which  occurs  in
the  BV  quantization  scheme.   Also,   one   should   note
that   $Q$   is   nilpotent   $Q^{2}( \cdots )  =  0$    and
that   $\chi^{ i }$  is  an   element   of   the   cotangent
bundle over instanton moduli space. Hence, all this together
implies that we should think of $Q$ as the  deRham  operator
on instanton moduli  space.  Thus,  as  $Q$  is  the  deRham
operator on instanton moduli space and  observables  of  the
theory are $Q$ cohomology classes, observables of the theory
correspond to elements in the cohomology ring over instanton
moduli  space.  Equivalently,  one  may  refer  to  them  as
elements  in  the  cohomology  ring  over   the   space   of
supersymmetric maps \StrommingerII. We prefer  to  think  of
them as  elements in  the  cohomology  ring  over  instanton
moduli space  as  it does  not  introduce any extra nomenclature.


\vskip .25in
\noindent {\it 2.2.4 Correlation Functions of $W_{\psi}$ }
\vskip .25in

Now, as we know what the observables of the theory  are, let
us examine the correlation functions  of  such  observables.
Consider a group of such observables  ${\cal O}_{a}$. As  we
know, ${\cal O}_{a}$ is a cohomology  element  on  instanton
moduli space ${\cal M}_{[X]}$ for  maps  of  homotopy  class
$[X]$. We will assume ${\cal O}_{a}$ is  a  form  of  degree
$q_{a}$. The correlation function of these observables taken
in the sector of maps with homotopy  class $[X]$ is,

\eqn
\CorrelationFunctionI{
	\langle
		\prod_{a} {\cal O}_{a}
	\rangle_{[X]} =
	\int_{B_{[X]}}
	DX     D\chi^{i} Db_{0}  DC_{0}
	Dd_{1} DE_{1}    \,\,
		e^{ -W_{\psi} }
		\prod_{a} {\cal O}_{a},
}

\noindent where $B_{[X]}$ is the set of maps  with  homotopy
class $[X]$. Now, the  ${\cal O}_{a}$  can  always  be  made
independent   of  the  fields\foot{  As  $b_{0}$,   $C_{0}$,
$d_1$, and $E_1$ are non-minimal fields, all  ${\cal O}_{a}$
can be made independent of these fields.} $b_{0}$,  $C_{0}$,
$d_1$, and $E_1$. So,  in integrating over $b_{0}$, $C_{0}$,
$d_1$, and  $E_{1}$  one  does  not  have  to  worry   about
the ${\cal O}_{a}$. Thus, in integrating over $b_0$,  $C_0$,
$d_1$, and $E_1$ one obtains,

\eqn
\CorrelationFunctionII{
	\langle
		\prod_{a} {\cal O}_{a}
	\rangle_{[X]} =
	\int_{{\cal M}_{[X]}} \,\,
		e^{ -S_{0} }
		\prod_{a} {\cal O}_{a},
}

\noindent where we have employed the fact that by way of our
$t  \rightarrow  \infty$  argument  the   path-integral   is
localized to instanton moduli space  ${\cal M}_{[X]}$.  Now,
as  the  ${\cal O}_{a}$ are elements in the cohomology  ring
of ${\cal M}_{[X]}$,  it must be that the product $\prod_{a}
{\cal O}_{a}$ is a top-form on ${\cal M}_{[X]}$ or the above
integral vanishes identically. Hence, to have a  non-trivial
integral one must have,

\eqn
\GhostNumberConstraintI{
	\sum_{a} q_{a} = Dim( {\cal M}_{[X]} ).
}

\noindent But, we  have  seen  that  $\chi^{i}$  defines  an
element of the cotangent space  of ${\cal M}_{[X]}$. So, the
dimension of  ${\cal M}_{[X]}$  is given by the dimension of
the space of $\chi^{i}$'s which  satisfy  the  equations  of
motion for $C_{0}$ and $E_{1}$. Let us denote the  dimension
of  this  space  as  $a_{[X]}$.  However,  looking  at   the
equations of motion for both $C_{0}$ and $E_{1}$  one  finds
that generically they  imply  $\chi^i$  vanishes identically
and thus $a_{[X]}= 0$. So, this implies $Dim({\cal M}_{[X]})
= 0$. Thus, the  moduli space ${\cal M}_{[X]}$ is  simply  a
set of discrete points.


\vskip .25in
\noindent {\it 2.2.5 Correlation Function Example }
\vskip .25in

Consider the evaluation map $e$  defined
by

\eqn
\EvaluationMap{
\eqalign{
e: {\cal M}_{[X]}\times\Sigma_{3} &\rightarrow X_{6} \cr
( \,\,\, X \,\,\, , \,\,\, p \,\, ) \, &\rightarrow X(p). \cr
}
}

\noindent By way of this evaluation  map  we  may  pull-back
elements of $H^{*}( X_{6} )$ to obtain  elements  of  $H^{*}
( {\cal M}_{[X]}  \times  \Sigma_{3} )$.   Furthermore,   by
restricting           such             elements           of
$H^{*}( {\cal M}_{[X]}   \times  \Sigma_{3} )$     to     an
arbitrary $p \in\Sigma_{3}$ one obtains an element of $H^{*}
({\cal M}_{[X]})$ and thus an observable of the  generalized
topological sigma model. Let us use this process to obtain a
set of observables  of  the  generalized  topological  sigma
model.

Consider a set of cohomology elements $Y_a$ on  $X_6$  where
$a = 1, \dots , s$ with $Y_a$ of degree $q_a$. Employing our
above construction  and a choice of points $p_a \in\Sigma_3$
we  may define a set of observables ${\cal O}_{Y_{a}}(p_{a})
\equiv e^{*}(Y_{a})(p_{a})$ corresponding to the  cohomology
elements $Y_{a}$ on $X_6$. Let  us  consider the computation
of the correlation function,

\eqn
\CorrelationFunctionIII{
	\langle
		\prod_{a} {\cal O}_{Y_{a}}(p_{a})
	\rangle_{[X]} =
	\int_{B_{[X]}}
	DX     D\chi^{i} Db_{0}  DC_{0}
	Dd_{1} DE_{1}    \,\,
		e^{ -W_{\psi} }
		\prod_{a} {\cal O}_{Y_{a}}(p_{a}).
}

\noindent From our above discussion we know that  $\prod_{a}
{\cal O}_{Y_{a}}(p_{a})$ must be a top-form  on  ${\cal M}_{
[X]}$.  However, as $Y_{a}$ is a form of degree  $q_{a}$  on
$X_{6}$, ${\cal O}_{Y_{a}}(p_{a})$ is a form of degree $q_a$
on ${\cal M}_{[X]}$.  Hence,  the  above  path  integral  is
non-zero if, and only if,

\eqn
\GhostNumberConstraintII{
	\sum_{a} q_{a} = Dim( {\cal M}_{[X]} ).
}

\noindent Now, by way of  our  previous  arguments, we  know
that  $Dim({\cal M}_{[X]}) = a_{[X]}= 0$. Hence, ${\cal M}_{
[X]}$ consists of a discrete set of  points.  This, in turn,
implies that $\sum_{a} q_{a} = 0$, and thus, $q_{a} = 0$ for
all $a$. So, in other words, each  ${\cal O}_{Y_{a}}(p_{a})$
is a zero-form on ${\cal M}_{[X]}$. As  $Dim({\cal M}_{[X]})
= 0$,    the        computation        of                the
correlation   function   \CorrelationFunctionIII\    reduces
to  a  sum  of  terms;  each term in the sum  corresponds to
the contribution of one point  in  ${\cal M}_{[X]}$  to  the
correlation function. So, we must compute  the  contribution
of an arbitrary point of ${\cal M}_{[X]}$ to the correlation
function.  In  doing  this, one  finds  that    each   point
contributes a factor  ${\cal N}_{[X]}$  dependent  upon  the
boson and fermion determinants  at  that point.  However, as
a  result  of  BRST  symmetry,  one  finds  \WittenXI\  that
${\cal N}_{[X]}= 1$. Thus, if we denote the number of points
in ${\cal M}_{[X]}$ as $\sharp{\cal M}_{[X]}$, then we find

\eqn
\CorrelationFunctionIV{
	\langle
		\prod_{a} {\cal O}_{Y_{a}}(p_{a})
	\rangle_{[X]} =
	\sharp{\cal M}_{[X]}
	e^{ -S_{0} }
	\prod_{a} {\cal O}_{Y_{a}}(p_{a}).
}

\noindent If we sum over homotopy classes, we obtain,

\eqn
\CorrelationFunctionV{
	\langle
		\prod_{a} {\cal O}_{Y_{a}}(p_{a})
	\rangle = \sum_{[X]}
	\sharp{\cal M}_{[X]}
	e^{ -S_{0} }
	\prod_{a} {\cal O}_{Y_{a}}(p_{a}).
}

\noindent Hence, we have computed the  correlation  function
in this particular case. Let us now see how this relates  to
the ``physics" of M-Theory on $X_{6}$.


\newsec{Physics of the Generalized Topological Sigma Model}

In this section we will examine the  implications  that  the
generalized topological sigma model has for the  physics  of
M-Theory  compactified  on  a  six-dimensional   Calabi-Yau.
First, however, we  will  examine  a  class  of  generalized
topological sigma model observables which  are  a  bit  more
general than those we have studied up until now.


\subsec{ Generalized Observables }

In  this  subsection  we  will  examine  a  new   class   of
generalized topological sigma model observables.

Consider the evaluation  map  $e$  we  employed  earlier  to
construct observables for the generalized topological  sigma
model. By way of $e$ we can  pull-back  elements  of  $H^{*}
(X_{6})$ to ${\cal M}_{[X]} \otimes \Sigma_{3}$. Earlier, we
simply restricted such a pull-back to a  $p \in \Sigma_{3}$.
However,  now  we will do something a bit more general

Consider a $p$-form $Y \in H^{p}(X_{6})$. By way of  $e$  we
may pull $Y$ back to ${\cal M}_{[X]} \otimes \Sigma_{3}$  to
obtain  a  $p$-form  $e^{*}(Y)$  on  ${\cal M}_{[X]} \otimes
\Sigma_3$. Now, consider a homology element $y_{H} \in H_{q}
(\Sigma_{3})$. In integrating $e^{*}(Y)$  over  $y_{H}$  one
obtains a $(p - q)$-form on ${\cal M}_{[X]}$; namely

\eqn
\GeneralizedObservableI{
	i_{*}( e^{*}(Y) ) =
	\int_{y_{H}}
		e^{*}(Y).
}

\noindent  Furthermore,  as  $\chi^{i}$  is  a  one-form  on
${\cal M}_{[X]}$ and has ghost number 1,  $i_{*}(e^{*}(Y))$,
being a $(p-q)$-form, has  ghost  number  $(p-q)$. Also,  as
$e^{*}(Y)$ defines  an  element  of  $H^{p} ( {\cal M}_{[X]}
\otimes \Sigma_{3})$,  so  $i_{*}( e^{*}(Y) )$  defines   an
element of $H^{p-q}({\cal M}_{[X]})$ and thus an  observable
of the generalized topological sigma model. So, what we have
is a family of observables.  One  can  take  a  $Y \in H^{p}
(X_{6})$ and obtain a  set  of  various  observables  $i_{*}
( e^{*}(Y) )$ dependent upon  the  $y_{H}$  one  chooses  to
integrate over.

In addition to this family of observables, we  also   obtain
a family of generalized topological sigma  models from  this
construction.  Consider   letting  $d_{I}$  be  a symplectic
basis for $H^{3}(X_{6})$.  If  we  choose   as  our homology
element $\Sigma_3$, then this leads to a set of observables,

\eqn
\GeneralizedObservableII{
	\int_{\Sigma_{3}}
		e^{*}(d_{I}).
}

\noindent With  these  observables  we  can  define  another
generalized topological sigma model  by  choosing  a  vector
$t^{I} \in {\bf R}^{b_{3}}$,  where  $b_{3}$  is  the  third
Betti number of $X_{6}$. This vector allows us to  define  a
perturbation of our original action given by,

\eqn
\GeneralizedActionI{
	W(t^{I}) =
	W +
	\left(
		{ {1} \over {L^{3}_{P}} }
	\right)	
	\int_{\Sigma_{3}}
		i t^{I} e^{*}(d_{I}).
}

\noindent One should note a few things  about  this  action.
First,  as  $d_{I}$  is  a   three-form   and   $\Sigma_{3}$
three-dimensional, $\int_{ \Sigma_{3} } e^{*}( d_{I} )$  has
ghost number 0. Thus, this family of generalized topological
sigma  models  conserves  ghost  number.  Also,  one   could
consider  an  even  more  general  family   of   generalized
topological sigma models by starting with a form  of  degree
greater than three.  However,  we  will  not  consider  such
theories  here.  Let us now employ all of this to  establish
a  connection   between   M-Theory   and   our   generalized
topological sigma model.


\subsec{ The M-Theory Connection }

In this subsection we will establish  a  connection  between
M-Theory and the generalized topological sigma model. As was
established  some  time  ago  \TownsendI,  M-Theory   on   a
six-dimensional Calabi-Yau has  a  term  in  the  low-energy
effective action  of the form,

\eqn
\LowEnergyActionTermI{
	\int d^{5}x \sqrt{ g } \, \,
		{\overline \psi}^{I} {\psi}^{J}
		{\overline \psi}^{K} {\psi}^{L}
		R_{IJKL},
}

\noindent where $\psi^I$ are fermions and  $R_{IJKL}$  is  a
Riemann  curvature  on  the  hypermultiplet   moduli   space
\StrommingerII. One then can compute the corrections to this
four-fermion coupling from a space-time   point-of-view   to
obtain a correction \StrommingerII,

\eqn
\FourFermionCorrectionI{
		e^{
			-S_{0}
		  }
		\int_{\Sigma_{3}} X^{*}(d_{I})
		\int_{\Sigma_{3}} X^{*}(d_{J})
		\int_{\Sigma_{3}} X^{*}(d_{K})
		\int_{\Sigma_{3}} X^{*}(d_{L}).
}

\noindent This is the correction for a given homotopy  class
$[X]$, represented by $X$, under the assumption  ${\cal M}_{
[X]}$ consists of a single point.

Now,  as  we  found  in  the  previous  subsection,  we  can
introduce  a  set  of   observables   of   the   generalized
topological sigma model which are related to the  symplectic
basis $d_{I}$ for $H^{3}(X_{6})$. They are,

\eqn
\GeneralizedObservableIII{
	{\cal O}_{I} \equiv 
	\int_{\Sigma_{3}}
		e^{*}(d_{I}).
}

\noindent However, looking at the definition of $e$  we  may
equivalently write these as,

\eqn
\GeneralizedObservableIV{
	{\cal O}_{I} \equiv
	\int_{\Sigma_{3}}
		X^{*}(d_{I}).
}

\noindent   In   comparing   this   to    the   world-volume
observable used in \StrommingerII\ to compute the correction
\FourFermionCorrectionI,  one finds  that  this  is both  an
observable for the low-energy effective world-volume  theory
and an observable  for  the  generalized  topological  sigma
model. Thus, looking at \StrommingerII\ we  can  see exactly
what we must  do,  consider  the  correlation  function   of
four such observables for a given homotopy class $[X]$,

\eqn
\FourFermionCorrectionII{
	\left<
		{\cal O}_{I}
		{\cal O}_{J}
		{\cal O}_{K}
		{\cal O}_{L}
	\right>_{[X]} =
	\int_{B_{[X]}}
	DX D\chi Db_0 DC_0 Dd_1 DE_1 \,\,
		e^{-W_{\psi}}
		{\cal O}_{I}
		{\cal O}_{J}
		{\cal O}_{K}
		{\cal O}_{L}.
}

\noindent Now, if this integral  is  non-trivial,  then  the
product of the ${\cal O}$'s must be a top-form on  instanton
moduli space. But, as $d_I$ is a three-form and $\Sigma_{3}$
three-dimensional,  each  ${\cal O}$  is  a   zero-form   on
instanton moduli space. However, as we have found $a_{[X]} =
Dim({\cal M}_{[X]})= 0$, the dimensions of  ${\cal M}_{[X]}$
and ${\cal O}_{I} \cdots {\cal O}_{L}$  match up. Thus, this
integral reduces to  a  sum  where  each  term  in  the  sum
corresponds to the contribution of one point in  ${\cal M}_{
[X]}$ to the integral. If there exists only a  single  point
in ${\cal M}_{[X]}$, as assumed in \StrommingerII, then this
reduces to

\eqn
\FourFermionCorrectionIV{
		\left<
		{\cal O}_{I}
		{\cal O}_{J}
		{\cal O}_{K}
		{\cal O}_{L}
		\right>_{[X]} =
		e^{
			-S_{0}
		  }
		\int_{\Sigma_{3}} X^{*}(d_{I})
		\int_{\Sigma_{3}} X^{*}(d_{J})
		\int_{\Sigma_{3}} X^{*}(d_{K})
		\int_{\Sigma_{3}} X^{*}(d_{L}).
}

\noindent This is exactly the result \FourFermionCorrectionI\
obtained by Stromminger et. al.  in  their  field  theoretic
computation \StrommingerII. So, we have found  that  we  can
compute   corrections  to  the  four  fermion  couplings  in
M-Theory on a six-dimensional Calabi-Yau by simply computing
correlation functions in this generalized topological  sigma
model. A relatively novel result. More  generally,  assuming
${\cal M}_{[X]}$ consists of $\sharp{\cal M}_{[X]}$ points,

\eqn
\FourFermionCorrectionV{
		\left<
		{\cal O}_{I}
		{\cal O}_{J}
		{\cal O}_{K}
		{\cal O}_{L}
		\right>_{[X]} =
		\sharp{\cal M}_{[X]} \,\,
		e^{
			-S_{0}
		  }
		\int_{\Sigma_{3}} X^{*}(d_{I})
		\int_{\Sigma_{3}} X^{*}(d_{J})
		\int_{\Sigma_{3}} X^{*}(d_{K})
		\int_{\Sigma_{3}} X^{*}(d_{L}).
}

\noindent Which implies,

\eqn
\FourFermionCorrectionVI{
		\left<
		{\cal O}_{I}
		{\cal O}_{J}
		{\cal O}_{K}
		{\cal O}_{L}
		\right> =
		\sum_{[X]}
		\sharp{\cal M}_{[X]} \,\,
		e^{
			-S_{0}
		  }
		\int_{\Sigma_{3}} X^{*}(d_{I})
		\int_{\Sigma_{3}} X^{*}(d_{J})
		\int_{\Sigma_{3}} X^{*}(d_{K})
		\int_{\Sigma_{3}} X^{*}(d_{L}).
}

\noindent This  is  the  generalization  of  the  result  of
Stromminger  et. al. \StrommingerII\ to the  case  in  which
${\cal M}_{[X]}$ has more than a single point.


\newsec{Conclusion}

In  this  article  we   have   constructed   a   generalized
topological sigma model and examined various  properties  of
this model. In particular, we found  that  this  model  will
allow  us  to  calculate  the  corrections  to  four-fermion
couplings in M-Theory on  a  six-dimensional  Calabi-Yau,  a
useful trick for a topological  field  theory.  However,  we
should note that we  have  not  specified  the  topology  of
$\Sigma_{3}$ in \FourFermionCorrectionVI.  If  the  standard
topological sigma model is any hint, then upon understanding
how to obtain the generalized topological sigma model by way
of a ``twist," we will understand what $\Sigma_3$ to  employ
above. More generally, we should note that it is possible to
relatively  easily construct similar generalized topological
sigma models  by examining the world-volume actions of other
$p$-branes. So, various fermion couplings may be computed in
a similar manner by constructing and  examining  the correct
generalized topological sigma model. It would be fruitful to
have a full  enumeration  of  such  generalized  topological
sigma models and the fermion couplings which they allow  one
to compute. We leave this as an exercise for the  interested
reader.

\listrefs
\bye